# Pyramids and cootie catchers: new massless fermions in 2D materials

Vladimir Damljanović[1], Radoš Gajić[1] and Igor Popov[1,2]
[1]Institute of Physics Belgrade, University of Belgrade, Pregrevica 118, Belgrade, Serbia
[2]Institute for Multidisciplinary Research, University of Belgrade, Kneza Višeslava 1a, Serbia

*Abstract*

Dirac-like electronic states are the main engines powering the tremendous advances in research of graphene, topological insulators and other materials with these states. Zero effective mass, high carrier mobility and numerous applications are some consequences of linear dispersion that distinguishes Dirac states. Here we report a new class of linear electronic bands in two-dimensional materials with zero effective mass and sharp band edges never seen in solid state matter before, and predict stable materials with such electronic structure utilizing symmetry group analysis and ab initio approach. We make a full classification of completely linear bands in two-dimensional materials and find that only two classes exist: Dirac fermions on one hand and pyramidal-like and cootie catcher-like states on the other hand. The new class supports zero effective mass and hence high carrier mobility similar to that of graphene, anisotropic electronic properties like that of phosphorene, and robustness of states with respect to electronic correlations.

*Introduction*

Electrons can move in certain materials as if they have no mass. Massless fermions in solid state materials have played an increasingly important role since the discovery of graphene (ref. [1]), a material where zero electron effective mass is caused by linear Dirac-like dispersion. While the first mapping of electronic structure of graphene to the Dirac equation was an interesting theoretical curiosity (ref. [2]), true significance of the Dirac-like states in solid state systems became apparent upon identification of many physical, measurable consequences of the linear dispersion (ref.[3]). For instance, the existence of massless fermions in graphene yields the extraordinary high electron and hole mobilities (ref.[4]), with revolutionary implications in electronics. Other implications include and are not limited to the Klein tunneling in single- and bi-layer graphene (ref. [5]), quantum Hall (ref. [6]) and fractional quantum Hall (ref. [7]) effects at room temperature. Two-dimensional (2D) nature of graphene and related materials bring numerous additional advantages including mechanical flexibility, optical transparency (ref. [8]), possibilities for engineering heterostructures with desired properties by stacking of two or more 2D materials (ref. [9]). There is a whole plethora of various 2D materials beyond graphene with the linear dispersion in their band structure (ref. [10]), but also in topological insulators (ref. [11]) and semimetals (ref. [12]). Dirac cones are intricately linked to symmetry, as exemplified with cone engineering by symmetry manipulation (ref. [13]), whereas new Dirac cones have been generated in graphene under external periodic potential (ref. [14]). The symmetry-electronic dispersion connection has been utilized in theoretical search of new materials with Dirac fermions. Mañes has used space group representation to find sufficient conditions for the existence of Weyl points (3D analogue of Dirac points) in the Brillouin zone (BZ) of three-dimensional single crystals (ref. [15]). Recently, a set of symmetry conditions that guarantees Dirac-like dispersion in the vicinity of high symmetry points in the BZ of any non-magnetic 2D material has been reported (refs. [16], [17]). Also, the existence of Dirac fermions in bilayer non-honeycomb crystals using symmetry analysis has been predicted (ref. [18]).

New band hourglass-like dispersion has been theoretically predicted in ref. [19], while fermionic excitations in electronic structure of three-dimensional materials have been theoretically classified in [20]. Surprisingly, the existence of other classes of massless fermions in 2D materials has not been addressed yet. Here we report that combined time-reversal (TRS) and certain crystal non-symmorphic symmetries of 2D materials lead to the emergence of peculiar massless linearly dispersive bands. The geometries of these states in reciprocal space are pyramidal-like and cootie catcher-like, which have never been seen in solid state matter before. Our analysis indicates that these states and the Dirac cones are the *unique* possibilities for non-accidental linear dispersive bands in all diperiodic directions of non-magnetic 2D materials without spin-orbit coupling. Finally, we predict a number of 2D materials with these massless fermions using density functional theory and our own-developed software.

*Classification of linear states in 2D*

The bands carrying zero effective mass are possible only in the vicinity of points in the Brillouin zone where the electron energy is (orbitally) degenerate. Energy dispersion of non-degenerate bands is smooth, hence second derivative is finite and its inverse, being proportional to the effective mass (ref. [21]), is nonzero. First we classify all possibilities for linear dispersions in the band structure of 2D materials. In order to achieve this aim we define a set of parameters, which values determine possible existence of linear dispersions in 2D crystals. If $G(\vec{k}_0)$ is the group of the wave vector $\vec{k}_0$ and $R$ is allowed (ref. [22]) (relevant (ref. [23]), small (ref. [24])) irreducible representation (irrep) of $G(\vec{k}_0)$, then the set of parameters consists of

➢ equivalence of $\vec{k}_0$ and its inverse $-\vec{k}_0$,
➢ dimensionality of representation $R$,
➢ reality of representation $R$.

Diperiodic groups have only 1D or 2D allowed irreps (ref. [25]), while they can be real on one hand or pseudo-real or complex on the other hand. When complex conjugation is a symmetry operation,

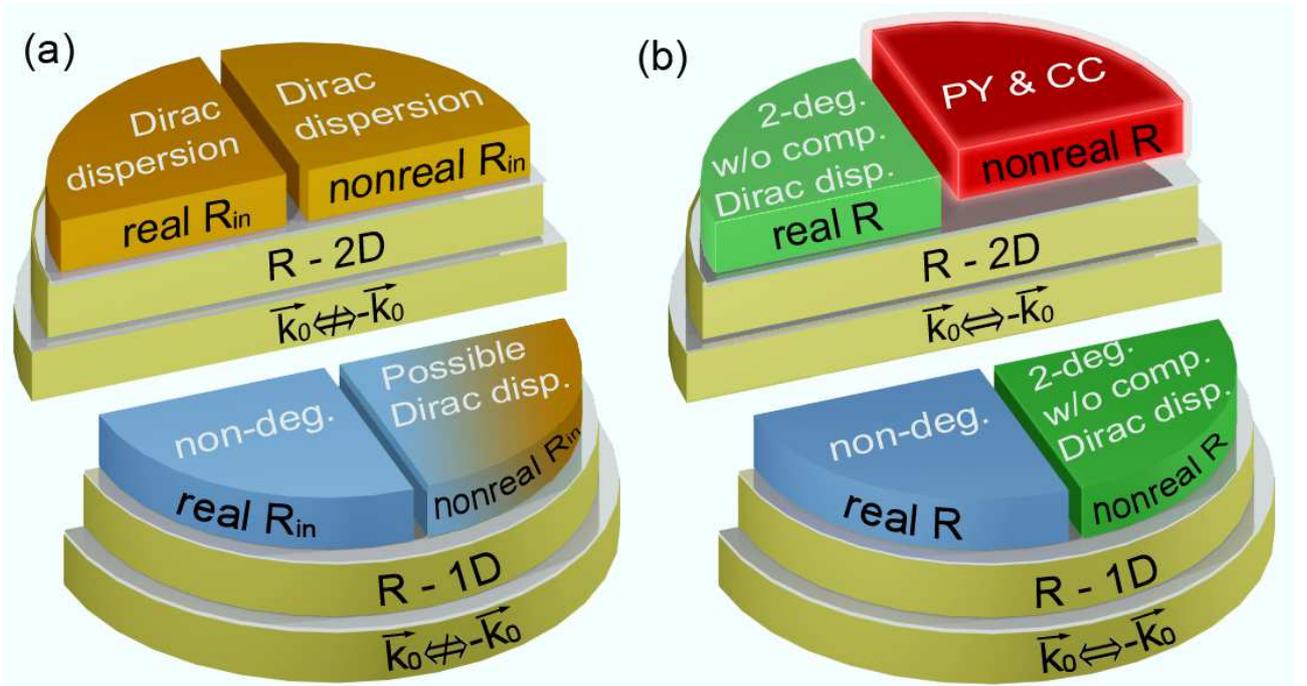

Figure 1 (color online) Full classification of linearly dispersive electronic bands in non-magnetic 2D materials based on symmetry conditions. Panel (a) corresponds to the case $\vec{k}_0 \not\Leftrightarrow -\vec{k}_0$ and panel (b) to the case $\vec{k}_0 \Leftrightarrow -\vec{k}_0$.

reality of irreps determines if it causes additional degeneracy. For single crystals the corresponding theory was developed in 1937 (ref. [26]). Therefore, each of these parameters can obtain one of two options; hence there are 8 possible combinations, as illustrated in Fig. 1.

We first consider the case $\vec{k}_0 \Leftrightarrow -\vec{k}_0$ (Fig. 1a). The wave vector $\vec{k}_0$ must have a locally maximal symmetry, otherwise linear dispersion cannot appear, due to either too many band contacts (ref. [27]) or none at all. If $R$ is two-dimensional, and the irrep $R_{in}$ of the whole diperiodic group $G$, which is obtained by induction from $R$ ($R_{in} = R\uparrow G$) is real, Dirac-like dispersion appears (ref. [16]) (orange upper left section). If $R_{in}$ is not real, $-\vec{k}_0$ is not in the star of $\vec{k}_0$ (ref. [22]) and the additional degeneracy due to TRS does not appear. This case also leads to Dirac dispersion (ref. [16]) (orange upper right section). In the last two cases double degeneracy at Dirac point is caused by the crystal symmetry. For $R$ one-dimensional and $R_{in}$ real (left panel blue section), $E_0$ is non-degenerate preventing Dirac dispersion in the vicinity of $\vec{k}_0$. For $R$ one-dimensional and $R_{in}$ pseudoreal or complex (blue-orange section) there are two possibilities. If $-\vec{k}_0$ is not in the star of $\vec{k}_0$, $E_0$ is non-degenerate while in the opposite case TRS causes $E_0$ to be double degenerate. In this case, there is a complete Dirac-like dispersion around $\vec{k}_0$ (ref. [17]). Next we consider the case $\vec{k}_0 \Leftrightarrow -\vec{k}_0$ (Fig. 1b). If $R$ is one-dimensional and real, the energy level $E_0$ at $\vec{k}_0$ is non-degenerate and linear dispersion cannot appear (right panel blue section). If $R$ is one-dimensional and not real (right panel green section down), TRS causes $E_0$ to be double degenerate, but complete linear Dirac-like dispersion is not possible since the TRS causes $u_2=0$ (ref. [16]). The same statement holds for two-dimensional, real $R$ (right panel green section up). The remaining case in which $R$ is two-dimensional and pseudoreal or complex will be treated in more detail here (right panel red section).

*Pyramidal and cootie catcher states*

In following, the functional form of the new linear dispersion relation that corresponds to the red section of Fig. 1 will be presented and compared to Dirac-like dispersion. In the following $\vec{q}$ is a wave vector of small modulus, $\vec{t}$ a real 2D vector, $u_1, u_2$ positive quantities and $q_1, q_2$ projections of $\vec{q}$ along mutually orthogonal directions. If $\vec{k}_0$ is a point that hosts a pair of Dirac cones, then the Taylor expansion of the electron energy around this point reads:

$$E_{1,2}(\vec{k}_0 + \vec{q}) \approx E_0 + \vec{t}\cdot\vec{q} \pm \sqrt{u_1 q_1^2 + u_2 q_2^2}. \qquad (1)$$

For $u_1 = u_2$ Dirac cones are isotropic and for $\vec{t} \neq 0$ the cones are tilted (ref. [28]). The new electronic dispersion presented in this paper is:

$$E_{1,2,3,4}(\vec{k}_0 + \vec{q}) \approx E_0 \pm |u_1|q_1| \pm u_2|q_2||. \qquad (2)$$

Note that energy level $E_0$ is double orbitally degenerate in (1) and 4-fold degenerate in (2). While the Dirac band has the geometric form of simple cone with circular or elliptical cross section, the geometry of dispersion (2) consists of two different geometric forms. The plus sign under the absolute value yields the geometry of a four-sided pyramid (PY), whereas the minus sign corresponds to a more complex geometry, which looks like the paper origami called cootie catcher (CC) (Fig. 3(b)).

Now we show briefly how the linear dispersion (2) is connected to TRS and crystal symmetry, whereas more details are given in supplementary materials. A general matrix form of Taylor expansion of Hamiltonian around a given $\vec{k}_0$ point of BZ of non-magnetic 2D material is $\widehat{H}(\vec{k}_0 + \vec{q}) \approx E_0 \hat{I}_4 + \widehat{H}'$, where $\widehat{H}' = \widehat{W}(\hat{I}_4 \otimes \vec{q})$, $\widehat{W} = \langle\frac{\partial}{\partial\vec{q}}|\widehat{H}(\vec{k}_0 + \vec{q})|_{\vec{q}=0}$, $\hat{I}_4$ is four-dimensional unit matrix, while $\widehat{W}$ is four-by-eight matrix. Diagonalization of such a Hamiltonian leads to a

**TABLE I. Diperiodic groups hosting the dispersion (2) in the vicinity of BZ corners.**

| Diperiodic group | | Corresponding space group | | | Diperiodic plane | $R$ |
|---|---|---|---|---|---|---|
| Dg33 | $p\ b\ 2_1\ a$ | 29 | $P\ c\ a\ 2_1$ | $C_{2v}^5$ | $y = 0$ | $U_1$ |
| Dg43 | $p\ 2/b\ 2_1/a\ 2/a$ | 54 | $P\ 2_1/c\ 2/c\ 2/a$ | $D_{2h}^8$ | $y = 0$ | $U_1, U_2 \Leftrightarrow U_1^*$ |
| Dg45 | $p\ 2_1/b\ 2_1/m\ 2/a$ | 57 | $P\ 2/b\ 2_1/c\ 2_1/m$ | $D_{2h}^{11}$ | $x = 0$ | $T_1, T_2 \Leftrightarrow T_1^*$ |

*Notes:* The notations for diperiodic and space groups are according to Kopsky and Litvin (ref. [29]) and Hahn (ref. [30]), respectively. The *x*-, *y*- and *z*-axes are along a-, b- and c-directions of the orthorhombic 3D unit cell, respectively. The notation for allowed representation in the last column is according to the Bilbao Crystallographic Server (ref. [22]). ⇔ denotes equivalence between representations.

general functional form of dispersion relation:

$$E_{1,2,3,4} = E_0 \pm \tag{3}$$

$$\pm \frac{1}{\sqrt{2}} \sqrt{\sum_{j=1}^{6}(\vec{v_j} \cdot \vec{q})^2 \pm \sqrt{\left[\sum_{j=1}^{6}(\vec{v_j} \cdot \vec{q})^2\right]^2 - 4[(\vec{v_1} \cdot \vec{q})(\vec{v_6} \cdot \vec{q}) - (\vec{v_2} \cdot \vec{q})(\vec{v_5} \cdot \vec{q}) + (\vec{v_3} \cdot \vec{q})(\vec{v_4} \cdot \vec{q})]^2}}$$

where $\vec{v_j}$ are real 2D vectors. Now we introduce conditions, which the allowed representation must satisfy in order to have PY and CC states (red section in Fig. 1).

- $O_1$: $\vec{k}_0$ is equivalent to its inverse $-\vec{k}_0$,
- $O_2$: $R$ is two-dimensional,
- $O_3$: $R$ is pseudo-real or complex.

It turns out that only 3 out of 80 diperiodic groups, Dg33, Dg43 and Dg45, have allowed representations satisfying the conditions $O_1$ - $O_3$. For all of them we have found the form that symmetry imposes on the vectors $\vec{v_j}$ using Wigner's method of group projectors. Upon their insertion into equation (3) it reduces to the linear dispersion (2) for the three groups.

The groups allowing the dispersion (2) are listed in the Table I. All three groups are non-symmorphic and belong to the rectangular system. The component $q_1$ can be chosen as projection of $\vec{q}$ along direction that is parallel to any screw axis $2_1$, $q_2$ is projection along the perpendicular direction. The points $\vec{k}_0$ hosting the dispersion (2) are located at the corners of the rectangle that presents the BZ border. The corresponding space groups from Table I denote the space groups that are obtained by periodic repetition of diperiodic groups' elements along the axis perpendicular to the diperiodic plane.

Due to the Bloch theorem, orbital wave functions must belong to an allowed irrep at a given point in the Brillouin zone. This statement is valid irrespectively of strength of electronic correlations, since the Coulomb repulsion between electrons has the same transformation properties as the rest of the Hamiltonian. Allowed irreps of groups listed in Table I are the only ones at these points of the BZ so the electronic correlations cannot change the dispersion (2).

Symmetry of the crystal lattice is responsible for (an)isotropy of single crystals (ref. [31]). For example, isotropy of the electric susceptibility tensor in silicon is caused by the cubic symmetry, while the in-plane isotropy of graphene is caused by the hexagonal symmetry. In our cases, crystal axes are maximally of the second order, irreps of corresponding point groups are all one-dimensional and the materials belonging to diperiodic groups listed in Table I are expected to be *anisotropic*.

**TABLE II (Meta)stable 2D crystals with Dg45 group and 4-fold Wyckoff multiplicity.**

| Element | $E_{at}$(eV/at) | b(Å) | c(Å) | Dg45 coordinates (Å) | $\Delta E_F$(eV) | $v_b$($10^6$m/s) | $v_c$($10^6$m/s) |
|---|---|---|---|---|---|---|---|
| B | -6.51 | 3.12 | 2.97 | (0.390 0.387 0.743) | -0.65 | 1.22 | 1.52 |
| C | | | | Not found a stable structure with 4-fold multiplicity of Wyckoff positions | | | |
| Si | -5.56 | 3.74 | 4.64 | (0.765 0.583 1.160) | -2.30 | 0.91 | 0.79 |
| **P** | **-6.03** | **3.22** | **5.24** | **(0.780 0.708 1.310)** | **0.00** | **1.08** | **0.40** |

*Notes:* $E_{at}$ is atomization energy, b and c are lattice parameters. Coordinates of only one atom are given for each element. Other coordinates can be obtained from Wyckoff positions. $\Delta E_F = E_{PY-CC} - E_F$ – energy difference between Fermi level and nearest PY & CC states. Group velocities $v_b$ and $v_c$ are calculated using $v_i = \frac{1}{\hbar}\frac{\partial E(k_1, k_2)}{\partial k_i}$, where index *i* corresponds to *b*, or *c* lattice directions.

The effective mass for dispersions (1) and (2) is zero, which can be simply shown from $\left[\hat{m}_{eff}^{-1}\right]_{jl} = \hbar^{-2}\left[\frac{\partial^2}{\partial q_j q_l}E(q_1, q_2)\right]\bigg|_{q_1=q_2=0}$. Details of derivation are included in the supplementary material.

*Ab initio search for realistic materials*

Next we report (meta)stable 2D materials, which are predicted using ab initio calculations. We have developed and utilized software that automatically searches materials with a given group, analyzes their stability and band structure. The outline of algorithm is listed in the supplementary materials. We limited our search only to four elements of the periodic table that are known to build stable 2D materials of various crystal symmetries. We expect that many more stable materials consisting of other elements will be found in further research. The stable and metastable structures are listed in Table II, together with their structural and electronic parameters. Note that symmetry arguments presented above do not determine position of PY and CC states on the energy scale. Positioning of PY & CC states eventually at $E_F$ is of ultimate importance for their future experimental confirmation by e.g. angle resolved photoemission spectroscopy, but also because many physical phenomena originate only from states around $E_F$. We have applied recently reported theory (refs. [32], [33]) to Dg33, Dg43 and Dg45 groups and found that the number of electrons in valence states of a material must be 8*n*+4 per unit cell, where *n* is a positive integer, in order to be a zero-gap semiconductor. This is a necessary condition for PY & CC states to touch exactly at $E_F$ and that no other bands cross the Fermi level. More specifically, elemental crystals with listed groups with 4-fold multiplicity of Wyckoff positions can be such zero-gap semiconductors only if they contain elements with an odd number of valence electrons, including IA, IIIA, VA, VIIA and some other groups of the periodic table. Another option, 8-fold multiplicity of Wyckoff positions, guarantees metallic systems. Groups Dg33 and Dg43 do not have Wyckoff positions with 4-fold multiplicity, which remain in the same group when singly occupied. Therefore we present in Table II the elemental crystals with 4 atoms per unit cell, which includes only group Dg45. Note that compounds of 2 or more elements with any of Dg33, Dg43 and Dg45 symmetries can obey the condition for touching of PY & CC states at $E_F$. Importantly we have found a (meta)stable structure with PY & CC contact points positioned exactly at the Fermi level: P with Dg45 symmetry group. The structure of P (Dg45) is shown in Fig. 2(a). It consists of zig-zag chains of P atoms placed alternately at two parallel planes. The elementary unit cell contains 4 atoms, 2 per plane, with lattice parameters *b* = 3.22 Å and *c* = 5.24 Å. The potential energy surface of P (Dg45) has multiple local minima (many allotropes have been recently predicted (ref. [34])) so there are numerous (meta)stable phases of phosphorus with different symmetries. One of them is the system at the local

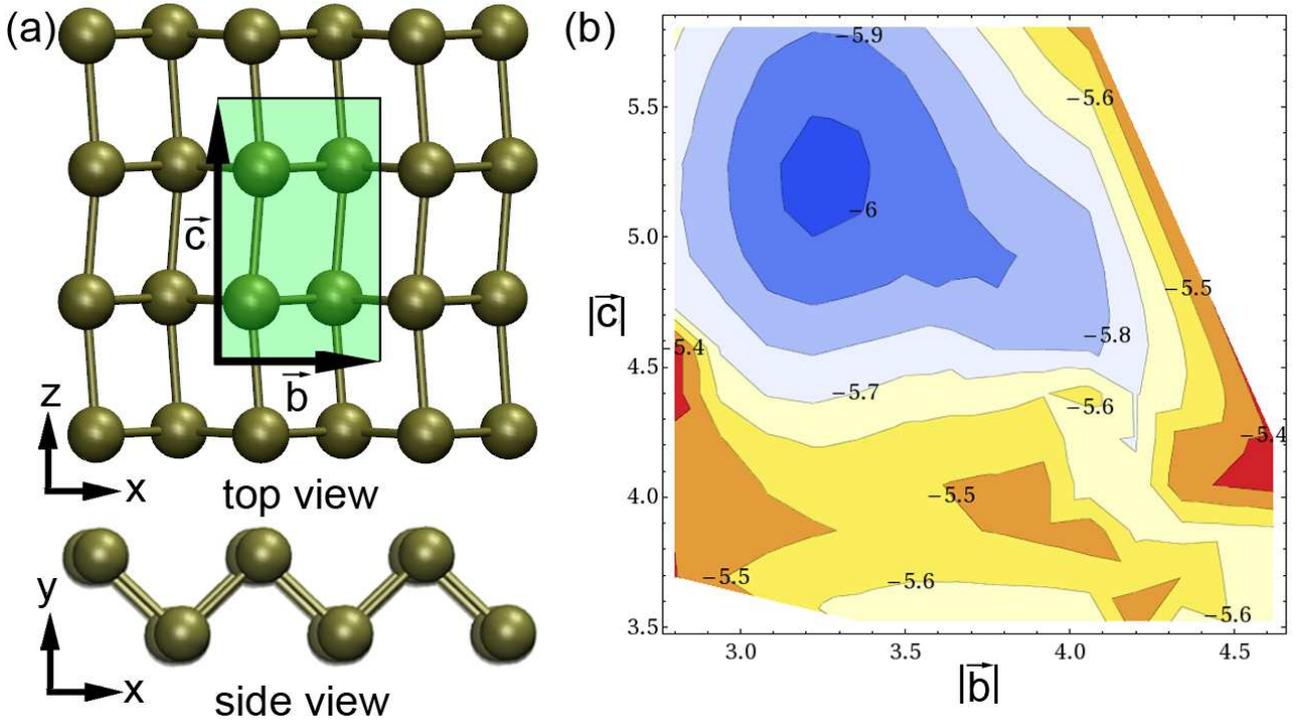

Figure 2 (color online) Top and side views of optimized geometry of P (Dg45) is shown in panel (a). An elementary unit cell is marked with a green rectangle together with lattice vectors. Potential energy surface (atomization energy given in eV/atom) with respect to lattice parameters is presented in panel (b). The lattice parameters are given in angstroms. Calculations are not done in white region of panel (b) since the search algorithm predetermined instability of the crystal in this region.

minimum shown in Fig. 2(b). Its stability is further confirmed by molecular dynamics simulations at 100K.

The band structure of P (Dg45) along lines between high symmetry points is shown in Fig. 3(a). At *X-R-Z* section of the BZ four states touch at $E_F$ (both upper and lower states are doubly degenerate), yielding 4-fold degeneracy at the point of contact ($E_F$). The bands around *R* point obey the electron-hole symmetry. The Fermi velocities of these states in *X-R* and *R-Z* directions are 1.08 and $0.46 \cdot 10^6$ m/s, which are in the range of the Fermi velocity of graphene. Therefore, we expect high electron and hole mobilities like that in graphene. The Fermi velocity of P (Dg45) is highly anisotropic; therefore anisotropic electronic properties are expected similar to those of phosphorene. The 4-fold degeneracy of bands at *R* is lifted along the diagonal direction (*Γ-R*). Note another set of bands between *Z* and *Γ*, with accidental 2-fold degeneracy below $E_F$. More geometry details of the states around *R* are visible in Fig 3(b). These states look exactly as the symmetry analysis predicted above: two pyramid-like and two cootie catcher-like bands touch at their tips and two lines, respectively. Due to the difference in Fermi velocities, coefficients $u_1$ and $u_2$ in eq. (2) are different, and the two lines intersect at an acute angle. Sharp edges of the dispersion are unique among electronic structure of any known crystal, and particularly in contrast to smooth features of Dirac cones.

In conclusion, we have established for the first time a full classification of states with linear dispersions in 2D materials based on group theory analysis, and found that only one additional class to Dirac states is possible. These states have not only unique and interesting geometric forms, but they can open new horizons for both fundamental research and applications. For instance, these fermions do not have an counterpart in elementary particle physics, i.e. they go beyond known Dirac, Mayorana and Weyl excitations. The sharp edges in electronic bands have been neither predicted nor measured before, so their existence in PY and CC states may spawn new phenomena in solid state materials. Possibly, very high mobilities on par with the mobility of graphene can be of uppermost interest for applications. Since the anisotropy of electronic properties widely opened the gates for phosphorene in the scientific community, materials with PY & CC bands can be

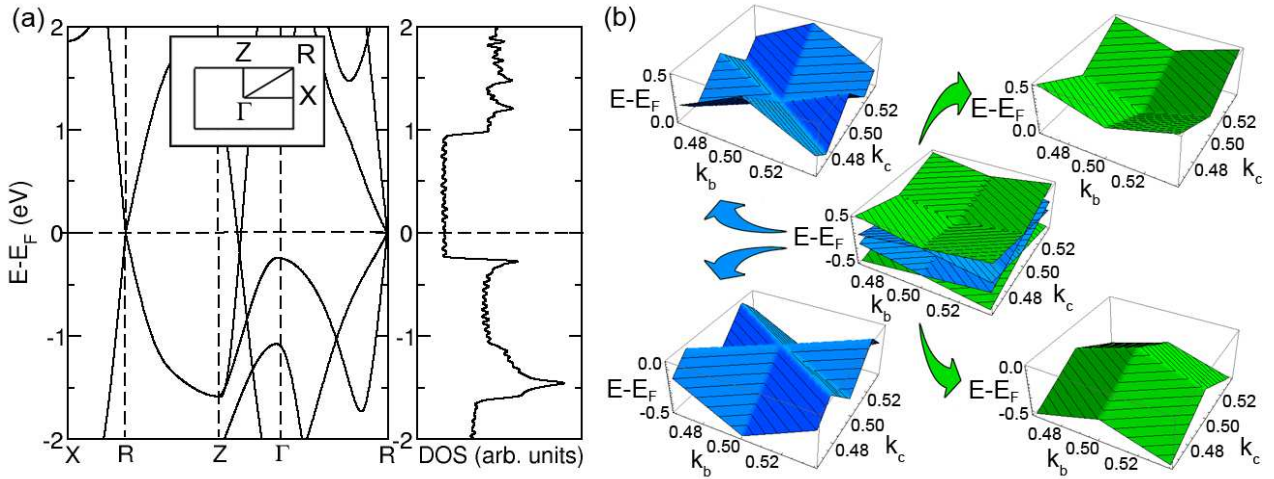

Figure 3 (color online) The band structure of P (Dg45). (a) Electronic band structure along lines between high-symmetry *k*-points (left panel) and corresponding density of states (right panel). (b) Pyramidal (green) and cootie catcher-like (blue) bands at point R. Energies are in units of eV relative to $E_F$, k points are in units of $2\pi/|\vec{b}|$ and $2\pi/|\vec{c}|$.

competitive to the popular phosphorene. The robustness of PY-PY and CC-CC contacts with respect to electron correlations may play a role in analogy to topological protection of states in topological insulators. Our unified classification of linearly dispersive bands paves the way to engineer new materials with Dirac and PY & CC states. We hope that findings presented here will be of large motivation for experimental groups to bring these materials into existence.

*Acknowledgements*
This work was supported by the Serbian Ministry of Education, Science and Technological Development under project numbers OI 171005 and III 45016. DFT calculations are performed using computational resources at Center of Surface and Nanoanalytics, Johannes Kepler University, Linz, Austria. We are grateful to Prof. Kurt Hingerl for providing necessary computational resources at JKU. We are thankful to Jelena Pešić and Marko Spasenović for discussions and suggestions related to this work.

*Author contributions*
VD performed group theoretical analysis and wrote a part of the manuscript. IP developed software, used it in combination with Siesta DFT package in the ab initio search of materials reported in this paper and wrote the most of the manuscript. RG gave the initial idea that the appearance of Dirac cones in the band structure of 2D materials is connected with their symmetry.

*Suplementary information*

More technical details of calculations used in this research are written in supplementary materials.

*Competing financial interests*

Authors declare no competing financial interests.

*Supplementary materials*

First we show how one obtains the dispersion relation (2) of the manuscript. Since $R$ is pseudoreal or complex, we can combine $R$ and its complex conjugate $R^*$ to obtain one four-dimensional, physically irreducible, real representation $D$ in the following way: $D=R+R^*$. We can choose the basis functions for $D$ to be real at $\vec{k}_0$, so that TRS imposes $\widehat{W}^* = -\widehat{W}$. Taking into account in addition the hermicity of the Hamiltonian, we get for $\widehat{H}'$:

$$\widehat{H}' = i\begin{pmatrix} 0 & \vec{v}_1 \cdot \vec{q} & \vec{v}_2 \cdot \vec{q} & \vec{v}_3 \cdot \vec{q} \\ -\vec{v}_1 \cdot \vec{q} & 0 & \vec{v}_4 \cdot \vec{q} & \vec{v}_5 \cdot \vec{q} \\ -\vec{v}_2 \cdot \vec{q} & -\vec{v}_4 \cdot \vec{q} & 0 & \vec{v}_6 \cdot \vec{q} \\ -\vec{v}_3 \cdot \vec{q} & -\vec{v}_5 \cdot \vec{q} & -\vec{v}_6 \cdot \vec{q} & 0 \end{pmatrix},$$

whose eigenvalues are given by the equation (3). Specific symmetry conditions of 2D crystals lead to the following relation which matrix $\widehat{W}$ obeys:

$$\left(\forall (\hat{h}|\vec{\tau}_{\hat{h}} + \vec{R}) \in G(\vec{k}_0)\right) \widehat{W} = \widehat{D}\left((\hat{h}|\vec{\tau}_{\hat{h}} + \vec{R})\right) \widehat{W} \left[\widehat{D}\left((\hat{h}|\vec{\tau}_{\hat{h}} + \vec{R})\right) \otimes \hat{h}'\right]^{\mathrm{T}},$$

where $\widehat{D}\left((\hat{h}|\vec{\tau}_{\hat{h}} + \vec{R})\right)$ is the matrix of the representation $D$ that corresponds to the element (in Seitz notation) $(\hat{h}|\vec{\tau}_{\hat{h}} + \vec{R})$ of the group of the wave vector $G(\vec{k}_0)$, and $\hat{h}'$ is the reduction (as an operator) of $\hat{h}$ to two-dimensional $\vec{k}$-space. It follows that the matrix $\widehat{W}$ belongs to the totally symmetric part of the representation $D \otimes D \otimes \Gamma_{2DPV}$, where $\Gamma_{2DPV}$ is two-dimensional polar-vector representation. Using Wigner's method of group projectors, we have found the form that the symmetry imposes on the vectors $\vec{v}_j$, and inserted the results in the equation (3). It turned out that in all cases, the result reduces to the dispersion (2).

Next we show that the dispersion (2) leads to zero effective mass. We take the dispersion for $E_1(\vec{q})$, obtained from (2) by taking plus signs, as an example. The proof for other bands, including Dirac-like (1) is analogous. Let us define the matrix:

$$\hat{S}(q_1, q_2) = \begin{pmatrix} \frac{\partial^2}{(\partial q_1)^2} E_1(q_1, 0) & \left[\frac{\partial^2}{\partial q_1 \partial q_2} E_1(q_1, q_2)\right]_{q_2=0} \\ \left[\frac{\partial^2}{\partial q_2 \partial q_1} E_1(q_1, q_2)\right]_{q_1=0} & \frac{\partial^2}{(\partial q_2)^2} E_1(0, q_2) \end{pmatrix}.$$

The matrix element $\left[\frac{\partial^2}{\partial q_1 \partial q_2} E_1(q_1, q_2)\right]_{q_2=0}$ means that we first take the derivative with respect to $q_2$, than take $q_2=0$, and then take derivative with respect to $q_1$ (and analogously for the other off-diagonal element). The matrix $\hat{S}$ in terms of Dirac delta function reads:

$$\hat{S}(q_1, q_2) = \begin{pmatrix} 2u_1 \delta(q_1) & 0 \\ 0 & 2u_2 \delta(q_2) \end{pmatrix}$$

Function $E_1(q_1, q_2)$ has minimum for $q_1 = q_2 = 0$, so the effective mass tensor is:

$$\widehat{m}_{\mathrm{eff}} = \hbar^2 \left[\hat{S}(q_1, q_2)\right]^{-1}\big|_{q_1=q_2=0}.$$

Using the interpretation of delta-function as being infinite at zero, we get $\widehat{m}_{\mathrm{eff}} = \hat{0}$.

Algorithm used in ab initio search

1. One chemical element is chosen from the main groups of the periodic table (IIIA, IVA and VA). Particularly we considered B, C, Si and P.
2. **Setup of initial parameters.** Initial fractional coordinates of a single atom, lattice vectors and one of diperiodic groups, Dg33, Dg43 or Dg45, are chosen. The initial lattice vectors and coordinates are chosen such that bond lengths in the system are as close as possible to typical bond lengths between atoms of given element. This is done manually. Set up the initial value of variable *current_minimal_energy* to a large value, i.e. 999999. Set up the initial *scaling factors* for lattice vectors. Value of 0.8 was a usual choice.
3. Scale lattice vectors by the current *scaling factors*.
4. Atomic positions of remaining atoms in a unit cell are generated based on Wyckoff equivalent positions for the chosen group.
5. **Screening of possibly stable geometries.** The generated crystal structure (combination of coordinates and scaled lattice vectors) is checked if it is *likely* stable by analyzing eventual clustering of its atomic positions to disjoined set of clusters. Structure is assumed as unstable and disregarded for further calculations if it has more than 2 disjoined clusters. Two clusters are considered disjoined if distance between two atoms closest to each other but belonging to different clusters is larger by more than $n$ Å than the sum of their covalent radii. Structure is also disregarded as unsuitable for further calculations if distance between clusters is smaller by $n$ Å than sum of their covalent radii. $n = 0.3$ Å was usually used for most of elements. Larger values of $n$ up to 0.5 Å are used for carbon, for which it is known to make a larger range of possible bond lengths. This step is done in order to speed up the search of stable structures.
6. **Symmetry-constrained geometry optimization.** If the given structure is not disregarded at the step 5 as an unsuitable initial geometry, a geometry optimization is conducted with constrained diperiodic group, i.e. its corresponding space group. Otherwise, go to the step 9. Density functional theory (DFT)-based software Siesta [1] is used for calculations of energies and atomic forces during the group-constrained geometry optimization.
7. **Can the symmetry be preserved?** Full unconstrained structural optimization is conducted using the Siesta code. Initial geometry for the optimization is the geometry obtained at the step 6. The optimized geometry is checked for eventual breaking of the symmetry after the full structural optimization. Also check the structural stability of the crystal using the same method described at the step 5.
8. If the symmetry is preserved and the structural stability is confirmed at the step 7, compare total energy with the *current minimal energy*. If it is a smaller one then promote it to the *current minimal energy*, and save atomic coordinates of this structure.
9. Increase scaling factors by 0.02. If they are smaller than 1.2 return to the step 3. Otherwise continue to the step 10.
10. Calculate electronic band structure for the most stable system (*current minimal*) using the Siesta code and analyze the band structure.

All ab initio calculations were done using DFT as implemented in Siesta code. Space groups from Table I were used, which correspond to diperiodic groups, when unit cells were constructed. Lattice vector perpendicular to diperiodic plane was always 15 Å. We utilized the Perdew-Burke-Ernzerhof form of the exchange-correlation functional [2]. The behavior of valence electrons was described by norm-conserving Troullier-Martins pseudopotentials [3]. We used a double-zeta polarized basis. The mesh cutoff energy of 250 Ry was used, which was sufficient to achieve a total energy convergence of better than 0.1 meV per unit cell during the self-consistency iterations of all calculations. Structures were considered as optimized when maximal force on atoms dropped below 0.04 eV/Å. In the search algorithm a 8 x 8 k-point Monkhorst-Pack mesh in plane of BZ corresponding to the plane of 2D materials was employed and only gamma point was used in perpendicular direction. A denser k-point mesh of 12 x 12 was used for further optimization and calculation of band structure and density of states of the most stable structures obtained by the search algorithm.
Bands in Fig. 3(b) were obtained on 300 x 300 k-point grid around the corner of BZ (point R).

Molecular dynamics simulation, which confirmed structural stability of P (Dg45) system, was conducted for 5 ps in 5000 steps of 1fs. Temperature was fixed at 100 K using the Nosé-Hoover thermostat. A super-cell comprising 3 x 3 x 1 repetition of a unit cell containing 36 P atoms was employed in the simulation.